\documentclass[apj]{emulateapj}
\usepackage{apjfonts}
\usepackage{float}
\usepackage{booktabs}
\usepackage{amsmath}

\shorttitle{Spatially-resolved Balmer decrements}
\shortauthors{Nelson et al.}
\slugcomment{Submitted to ApJ Letters}

\begin{document}

\gdef\ha{H$\alpha$}
\gdef\hb{H$\beta$}
\gdef\nii{[N\,{\sc\,ii}]}
\gdef\sii{[S\,{\sc\,ii}]}
\gdef\oiii{[O\,{\sc\,iii}]}
\gdef\m{M$_*$}
\gdef\msun{M$_{\odot}$}
\gdef\f{$H_{F140W}$}
\gdef\a{$A_{\rm H\alpha}$}
\gdef\av{$A_{v}$}
\gdef\iruv{IR/UV}
\gdef\kha{$k(\lambda_{H\alpha})$}
\gdef\khb{$k(\lambda_{H\beta})$}

\title{Spatially-resolved dust maps from Balmer decrements in galaxies 
at $z\sim1.4$}

\author{Erica June Nelson\altaffilmark{1}, 
Pieter G. van Dokkum\altaffilmark{1},
Ivelina G. Momcheva\altaffilmark{2},
Gabriel B. Brammer\altaffilmark{2},
Stijn Wuyts\altaffilmark{3},
Marijn Franx\altaffilmark{4},
Natascha M. F\"orster Schreiber\altaffilmark{5},
Katherine E. Whitaker\altaffilmark{6}\altaffilmark{$\dagger$},
Rosalind E. Skelton\altaffilmark{7}
}

\altaffiltext{1}{Astronomy Department, Yale University, 
New Haven, CT 06511, USA}
\altaffiltext{2}{Space Telescope Science Institute, 3700 San Martin Drive,
Baltimore, MD 21218, USA}
\altaffiltext{3}{Department of Physics, University of Bath, 
Claverton Down, Bath, BA2 7AY, UK}
\altaffiltext{4}{Leiden Observatory, Leiden University, Leiden, The
Netherlands}
\altaffiltext{5}{Max-Planck-Institut f\"ur extraterrestrische Physik,
Giessenbachstrasse, D-85748 Garching, Germany}
\altaffiltext{6}{Department of Astronomy, University of Massachusetts,
Amherst, MA 01003, USA}
\altaffiltext{7}{South African Astronomical Observatory, P.O.\ Box 9,
Observatory, 7935, South Africa}
\altaffiltext{$\dagger$}{Hubble Fellow}

\begin{abstract}
We derive average radial gradients in the dust attenuation towards H\,{\sc ii}
regions in 609 galaxies at $z\sim 1.4$, using measurements of the
Balmer decrement out to $r\sim 3$\,kpc. 
The Balmer decrements are derived from spatially resolved maps
of \ha\ and \hb\ emission from the 3D-HST survey.
We find that with increasing stellar mass
\m\ both the normalization and strength of 
the gradient in dust attenuation increases.
Galaxies with a mean mass of $\langle \log M \rangle = 9.2$\,M$_{\odot}$
have little dust attenuation at all radii,
whereas galaxies with $\langle \log M \rangle= 10.2$\,M$_{\odot}$ have
$A_{{\rm H}\alpha} \approx 2$\,mag in their central regions.
We parameterize this as
$A_{{\rm H}\alpha} = b + c\log r$, with $b = 0.9 + 1.0\log M_{10}$,
$c = -1.9 - 2.2\log M_{10}$, $r$ in kpc, and $M_{10}$ the
stellar mass in units of $10^{10}$\,M$_{\odot}$.
This expression can be used to
correct spatially resolved measurements of \ha\
to radial distributions of star formation. When applied to our data,
we find that the star formation rates in the central $r<1$\,kpc
of galaxies in the highest mass bin are $\sim 6$\,M$_{\odot}$\,yr$^{-1}$,
six times higher than before correction
and approximately half of the total star formation rate of these galaxies.
If this high central star formation rate is maintained for several Gyr,
a large fraction of the stars in
present-day bulges likely formed in-situ.
\end{abstract}

\keywords{galaxies: dust attenuation --- galaxies: evolution --- 
galaxies: structure ---  galaxies: star formation}.

\section{Introduction}

In the local Universe, the star formation surface density is determined
by the molecular gas surface density 
(e.g. {Larson} 1992).  
Recent studies have shown that this relation probably holds 
at higher redshift ($1<z<3$) as well
(e.g. {Tacconi} {et~al.} 2013; {Genzel} {et~al.} 2015), 
suggesting that the gas
reservoir is the key factor controlling the growth rate of galaxies.
In the broader context of galaxy evolution, this has been interpreted as
evidence 
for the equilibrium growth framework in which star formation is regulated by 
a balance between inflows and outflows 
({Bouch{\'e}} {et~al.} 2010; {Dav{\'e}}, {Finlator}, \&  {Oppenheimer} 2012; {Lilly} {et~al.} 2013; {Peng} \& {Maiolino} 2014).
In this model, galaxy growth in an integrated sense is driven by the cycle 
of baryons between galaxies and the halo gas ({Dav{\'e}} {et~al.} 2012) 
with the star formation efficiency set by stellar feedback ({Hopkins} {et~al.} 2014). 

However, in a resolved sense, exactly how different physical processes 
couple in different parts of galaxies to regulate star formation and drive  
structural assembly remains unknown ({Dav{\'e}} {et~al.} 2012).
Observationally, significant progress is being made in mapping the
molecular gas reservoir and the spatial distribution of star formation
of galaxies. 
Molecular gas can now be mapped based on the spatial distribution of 
CO thanks to millimeter arrays such as PdB and ALMA 
(e.g. {Genzel} {et~al.} 2013; {Rybak} {et~al.} 2015). 
Star formation can be mapped out to $z\sim 2.5$ using the spatial distribution
of the \ha\ emission line owing to integral field units on ground-based 
telescopes (SINFONI, OSIRIS, KMOS {F{\"o}rster Schreiber} {et~al.} 2009;
 {Law} {et~al.} 2009; {Wisnioski} {et~al.} 2015) 
and the WFC3 grism on HST ({Nelson} {et~al.} 2012; Nelson {et~al.} 2013).

However, studies of spatially resolved star formation using the \ha\ 
emission line suffer from an important systematic uncertainty
as they only trace the photons that are not absorbed by dust.
The dust attenuation toward star-forming regions
is most directly probed
using Balmer recombination line flux ratios, the Balmer decrement:
as dust attenuation is wavelength dependent, its effects can be
measured by comparing
the observed and intrinsic Balmer decrements 
(e.g. {Calzetti} 1997). 
On a galaxy-integrated basis,
the quantity of dust attenuation toward 
HII regions (\a) measured using
Balmer decrements has been shown to increase with increasing stellar mass (\m), 
star formation rate (SFR), and attenuation toward the stars (\av) 
({Calzetti} {et~al.} 2000; {Wild} {et~al.} 2011; {Momcheva} {et~al.} 2013; {Dom{\'{\i}}nguez} {et~al.} 2013; {Price} {et~al.} 2014; {Reddy} {et~al.} 2015). 
As the stellar mass surface density
and the star formation surface density are typically
highest in the centers of galaxies, we expect the attenuation
to vary within galaxies, such that
the dust attenuation is highest in the centers.
This is also expected from the fact that both the metallicity
and, most importantly, the gas column density increase with decreasing
distance from the center (e.g. {Bohlin}, {Savage}, \& {Drake} 1978; {Gilli} {et~al.} 2014; {Nelson} {et~al.} 2014).
Therefore, in order to tie the global gas reservoir
to galactic structural assembly we need to correct the
spatially-resolved \ha\
measurements for the attenuation toward star-forming regions.
 
Measuring spatially-resolved Balmer decrements
is now possible with the Wide Field Camera 3 (WFC3) grism 
capability on the Hubble Space Telescope, which we employed for 
the 3D-HST survey ({Brammer} {et~al.} 2012; {Momcheva} {et~al.} 2015).
The grism data enable us
to create emission line maps for every object in the survey
(see {Nelson} {et~al.} 2012; Nelson {et~al.} 2013; {Wuyts} {et~al.} 2013; {Nelson} {et~al.} 2015).
In a narrow redshift window ($1.35<z<1.5$)
we can map the spatial distribution of both the \ha\ and \hb\ emission 
lines, as they both fall
within the G141 wavelength coverage.
Galaxy-integrated Balmer
decrements were analyzed in {Price} {et~al.} (2014).

Here we present spatially resolved Balmer decrements for galaxies
at $z\sim 1.4$ and derive radial dust gradients. The gradients are
measured from deep stacks, using the full 3D-HST dataset.
We study radial dust gradients as a 
function of \m, and apply these gradients to the observed \ha\ distributions
to obtain spatially-resolved dust-corrected star formation surface density profiles.

\section{Spatially-Resolved Balmer Decrements}

We use data from the 3D-HST survey,
a 248 orbit NIR slitless spectroscopic survey over the 
CANDELS fields with the G141 grism ({Brammer} {et~al.} 2012; {Skelton} {et~al.} 2014; {Momcheva} {et~al.} 2015).
These slitless grism observations have high spatial resolution and
low spectral resolution, and therefore
provide images of galaxies in the light of their emission lines
for every object in the field of view. We focus on the redshift range
$1.35<z<1.5$, for which
both \ha\ and \hb\ fall in the wavelength coverage of the G141 grism.
The absence of sky emission lines in the spectra implies that
no further restrictions on the redshifts are required; in ground-based
observations it is rare that both \ha\ and \hb\ are unaffected by OH
lines. The galaxies are divided in three mass bins: 
[$9.0\leq M_* < 9.2$], [$9.2\leq M_* < 9.8$], [$9.8\leq M_* < 11.0$], 
The median stellar mass in these bins is 9.17, 9.53, and 10.23
respectively. 

\begin{figure}[hbtf]
\includegraphics[width=0.5\textwidth]{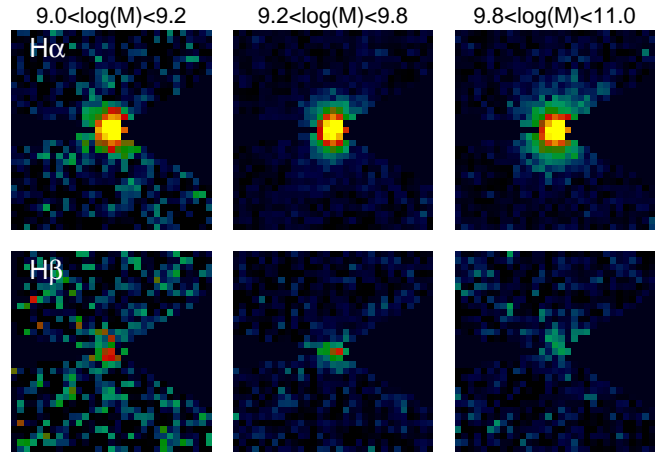}
\caption{Averaged maps of \ha\ and \hb\ emission,
in three different stellar mass bins.
These maps were obtained by stacking continuum-subtracted two-dimensional
spectra of galaxies at $1.35 \leq z \leq 1.53$ from the 3D-HST survey.
\label{fig:maps}
}
\end{figure}

\renewcommand{\thefootnote}{\fnsymbol{footnote}}
A detailed description of how emission line maps are made from grism data 
is provided in {Nelson} {et~al.} (2015).
Briefly, the \ha\ and \hb\ emission 
line maps are made by subtracting the continuum from the 
two-dimensional spectra and masking contaminating flux from nearby objects. 
We stack the \ha\ and \hb\ emission line maps as a function of 
\m.
These properties were determined from the combination of the 
grism spectra and deep UV-IR photometric catalogs 
({Brammer} {et~al.} 2012; {Skelton} {et~al.} 2014; {Whitaker} {et~al.} 2014). 
We select all galaxies with \f$\leq24$, applying 
no emission line flux limit for \ha\ or \hb\ so as not to 
introduce systematics into the line ratio measurements. 
Galaxies which have \ha\ or \hb\ emission line maps with 
more than half of their central 3\,kpc masked due to contamination
are removed; this is roughly half of the sample. 
In the stacking procedure
the broadband \f\ emission is used to center and normalize
the emission line maps. We masked the nearby
\sii\ $\lambda\lambda 6716,6731$\AA\ and 
\oiii\ $\lambda\lambda 4959,5007$ lines with an asymmetric double
pacman mask (see {Nelson} {et~al.} 2015). 
We correct the image stacks for the effects of the point spread function
(PSF), using the method described in {Szomoru} {et~al.} (2013) and
{Nelson} {et~al.} (2015): after fitting the stacks with GALFIT (Peng {et~al.} 2010),
the residuals of the fit are added
to the unconvolved GALFIT S\'{e}rsic model.\footnote[3]{Different PSFs were
created for the average observed wavelengths of the \ha\ and the \hb\ lines.}
This method has been 
shown to reconstruct the true flux distribution even if the S\'{e}rsic model
is a poor fit ({Szomoru} {et~al.} 2013).
The averaged maps are shown in Fig.\ \ref{fig:maps}.
Radial profiles of the \ha\ and \hb\ emission are computed in circular
apertures, again following {Nelson} {et~al.} (2015).

\begin{figure*}[hbtf]
\includegraphics[width=\textwidth]{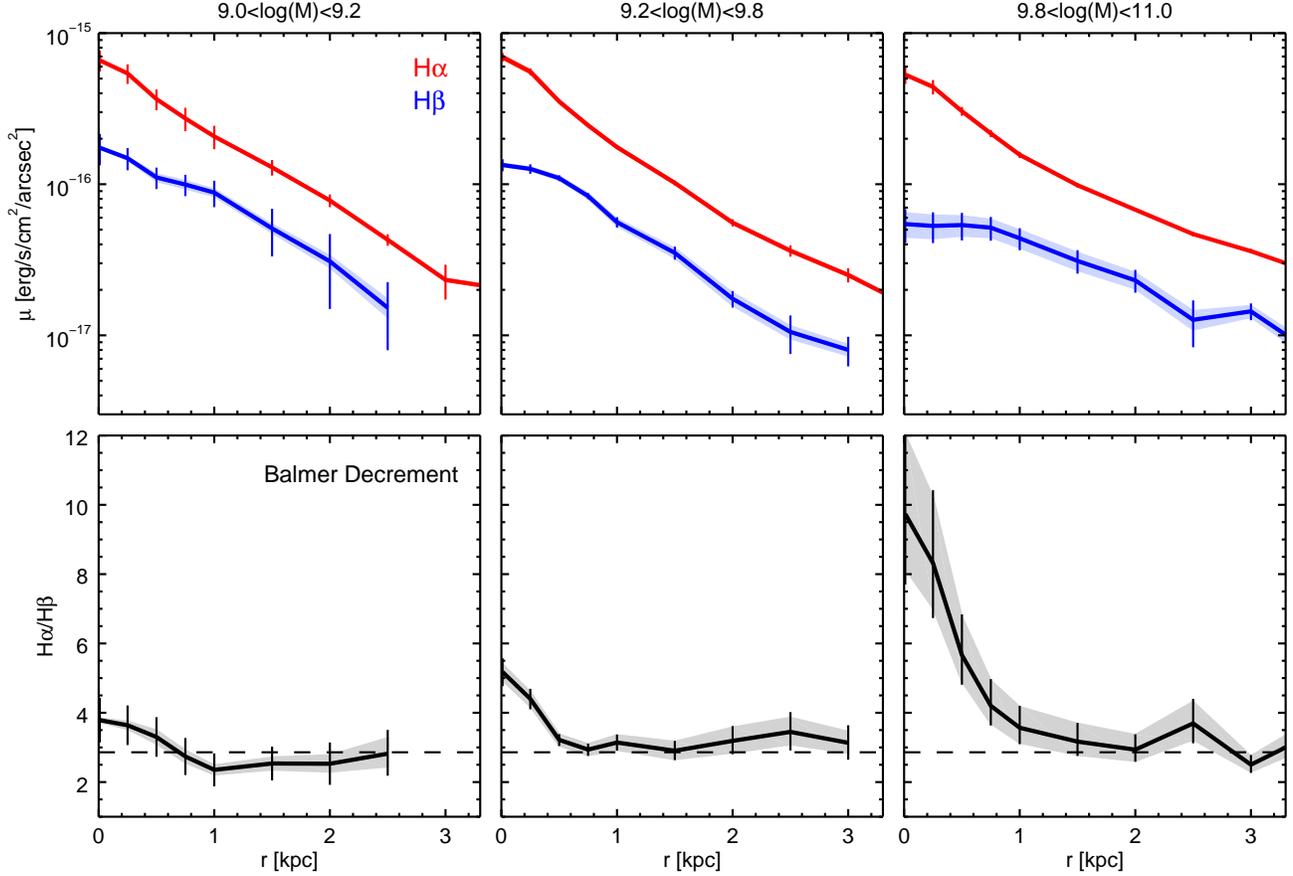}
\caption{Average radial surface brightness profiles of \ha\ 
(red), \hb\ (blue), and \ha/\hb\ (the Balmer decrement, black) 
in galaxies as a function of \m.  
Random uncertainties are shown by bootstrap error bars.
Systematic uncertainties derived by artificially increasing 
and decreasing the absorption line equivalent widths by half
are shown by shaded regions. 
The dashed line shows (\ha/\hb)$_{\rm int}$, the `intrinsic'
line ratio in the absence of dust attenuation.  
We can reliably measure 
the average Balmer decrement gradients in galaxies 
at $z\sim1.4$ to nearly 3\,kpc. With the Balmer decrement
tracing dust attenuation toward H\,{\sc ii} regions, this figure
shows that with increasing stellar mass, galaxies become
increasingly dust-obscured toward their centers.
\label{fig:habprofs}
}
\end{figure*}

\vspace{0.1cm}

We correct these measured profiles for \nii\ emission.
As a result of the low spectral resolution of the G141 grism,
\nii\,$\lambda 6548,6583$ and \ha\,$\lambda 6563$
are blended in our spectra. To account 
for the contamination of \ha\ by \nii, the blended observed line must 
be scaled by a factor \ha$_{\rm corr} =$\ha$_{\rm meas} / (1+$\nii/\ha).
We cannot measure \nii/\ha\ directly, and we make use of previous
measurements in the literature.
The galaxy-integrated \nii/\ha\ increases as a function 
of \m\ (e.g. {Erb} {et~al.} 2006; {Zahid} {et~al.} 2014; {Steidel} {et~al.} 2014; {Wuyts} {et~al.} 2014), 
which is probably a reflection of the mass-metallicity relation.
We use
the {Zahid} {et~al.} (2014) relation and scale the overall normalization of 
the radial \ha\ profiles down accordingly. The adopted \nii/\ha\ ratios are
[0.03,0.07,0.24] for the low through high mass stacks respectively. 

We also consider the effects of gradients in \nii/\ha\ within
the galaxies, as abundance gradients 
have been widely observed in 
the local Universe (e.g. {S{\'a}nchez} {et~al.} 2014).
However, trends at higher redshift are much less certain with observed 
\nii/\ha\ gradients ranging from flat or negative 
to positive (Swinbank {et~al.} 2012; {Jones} {et~al.} 2013; 
{Stott} {et~al.} 2014,{Leethochawalit} {et~al.} 2015, Wuyts et al. in prep). 
Given these uncertainties
we do not apply a radially-varying correction, but we note that
adopting a radial gradient of \nii/\ha$=-0.18\,{\rm dex}/r_e$,
as inferred from radial O/H measurements in the local Universe
({S{\'a}nchez} {et~al.} 2014), does not change our results.

Typically, Balmer emission lines also need to be corrected for underlying 
absorption. The atmospheres of stars (particularly A stars) produce Balmer
absorption lines and the emission line flux must first fill in the absorption. 
In our analysis, the absorption was corrected for in the subtraction of 
the two-dimensional continuum model. 
This model is a linear combination of EAzY templates with emission 
lines removed (see {Brammer}, {van Dokkum}, \&  {Coppi} 2008; {Momcheva} {et~al.} 2015) which is then convolved
with the $J_{F125W}/H_{F140W}/H_{F160W}$ detection image.
Therefore, the continuum model has, to very good approximation, 
both the same spectrum and morphology as the true continuum
emission that underlies our emission line maps.
We are essentially subtracting a negative image of the galaxy in 
absorption. 
For reference, the integrated absorption line strength is
$\sim 2-3.5$\,\AA\ in the stacks,
consistent with, e.g., {Dom{\'{\i}}nguez} {et~al.} (2013); {Momcheva} {et~al.} (2013). 
This is, in general, a small fraction of the emission line equivalent width,
although it becomes more significant for \hb\ at the highest masses 
(up to $\sim30\%$). 

The fully corrected \ha\ and \hb\ profiles, in units of 
$\textrm{erg}\,\textrm{s}^{-1}\,\textrm{cm}^{-2}\,\textrm{arcsec}^{-2}$,
 are shown in Fig.\,\ref{fig:habprofs}. We can reliably trace \ha\ out to 
$\gtrsim6\,{\rm kpc}$, and the $\gtrsim3\times$
fainter \hb\ out to $\sim 3\,{\rm kpc}$; at larger radii the error in
the measurement is more than half of the measured flux. 
At low masses, the \ha\ and \hb\ surface brightness
profiles are nearly exponential. 
As mass increases, the \ha\ emission grows  more
centrally concentrated while the \hb\  becomes {\em less}
centrally concentrated
than exponential. This is the central result of this Letter:
with increasing stellar mass, galaxies become increasingly more dust
obscured toward their centers.

Fig.\ \ref{fig:habprofs} also shows the effect of assuming different 
quantities of stellar absorption (shaded regions). These profiles
were derived by artificially changing the absorption line
equivalent width in the best-fitting 2D EAzY model, increasing and decreasing it 
by half.


\begin{figure*}[hbtf]
\includegraphics[width=\textwidth]{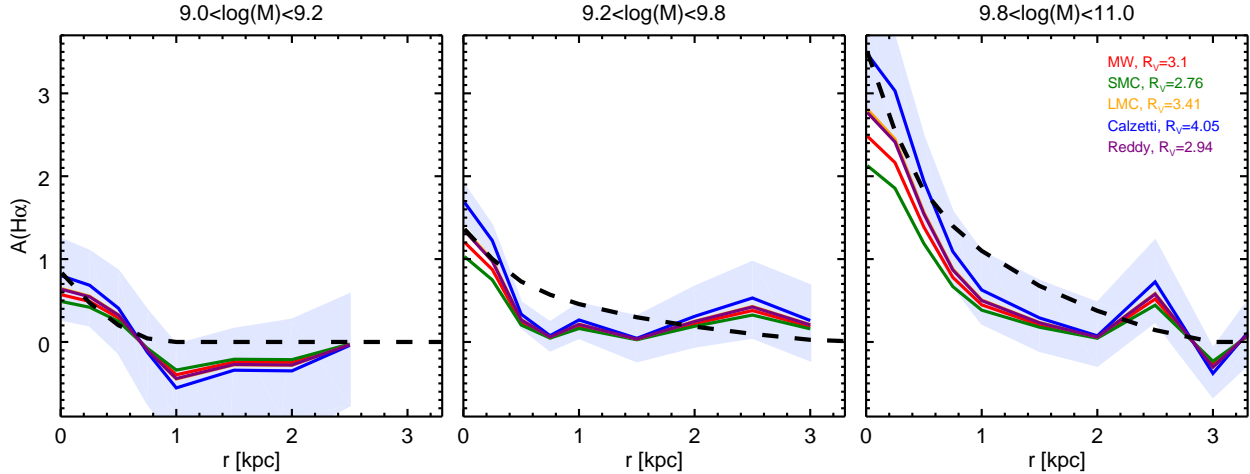}
\caption{Radial profiles of dust attenuation toward the \ha\ emission line
(\a) in galaxies as a function of \m. 
\a\ is computed by scaling the Balmer decrement using a dust law. 
Different dust laws result in different conversions from the radial Balmer decrement
profiles in the previous figure to the \a\ profiles here, shown in different colors.
({Calzetti} {et~al.} 2000; {Gordon} {et~al.} 2003; {Reddy} {et~al.} 2015). 
In summary, the star formation in the centers of galaxies becomes more
obscured with increasing stellar mass. 
\label{fig:ahaprofs}
}
\end{figure*}

\section{Radial Gradients in Dust attenuation}
The increase in the slope and normalization of the Balmer decrement
with \m\ implies a corresponding increase in the slope and 
normalization of the dust attenuation. We derive the dust attenuation 
toward \ha\ as follows. 
The increase of the Balmer decrement over the intrinsic value can be 
expressed in terms of a Balmer color excess:
\begin{equation}
E(H\beta-H\alpha)=2.5\,{\rm log}\left(\frac{(H\alpha/H\beta)_{obs}}
{(H\alpha/H\beta)_{int}}\right)
\end{equation}
Here (\ha/\hb)$_{\rm int}$ is the expected ratio produced
by the recombination and subsequent cascade of electrons
in photoionized hydrogen atoms: (\ha/\hb)$_{\rm int}=2.86$
for Case B recombination and 
$T=10^4{\rm K}$ ({Osterbrock} \& {Ferland} 2006). 

If dust is present, it will preferentially attenuate the shorter wavelength 
\hb\ $\lambda 4861$\AA\ more than the longer wavelength 
\ha\ $\lambda 6563$\AA, increasing the observed value of the 
Balmer decrement ((\ha/\hb)$_{\rm obs}$). 
The Balmer color excess can therefore
be used to derive the attenuation toward \ha:
\begin{equation}
A_{H\alpha}=\frac{E(H\beta-H\alpha)}
{k(\lambda_{H\beta})-k(\lambda_{H\alpha})}
\times k(\lambda_{H\alpha})
\end{equation}\\
Deriving \a\ requires the adoption of a reddening curve $k(\lambda)$
to compute the values at \ha\ (\kha) and 
\hb\ (\khb). 

In Fig.\ \ref{fig:ahaprofs} we
show the radial trend of the absorption toward \ha, \a\,
using different parameterizations of the reddening
and attenuation.
As expected from the trends in the Balmer decrement profiles, 
\a\ shows a stronger dependence on
radius and a higher overall
normalization for higher masses.
The attenuation in low mass galaxies is essentially zero at all radii.
As mass increases, attenuation increases toward the center. In the highest
mass bin, the attenuation in the central
regions reaches values in excess of two magnitudes.

While the qualitative trends remain unchanged with different
dust laws, the profiles do have quantitative differences.
In particular, \a\ is higher if the slope of the 
attenuation curve is flatter between the wavelengths of \ha\ and \hb\
or if $R_V$ is larger (see e.g. {Cardelli}, {Clayton}, \&  {Mathis} 1989; {Calzetti} {et~al.} 2000; {Gordon} {et~al.} 2003; {Reddy} {et~al.} 2015).
With the exception of the centers of the highest
mass galaxies, these
differences are small compared to the systematic and
random errors in the measurements of the Balmer decrement.

We quantify the radial dust profiles as follows:
\begin{equation}
\label{fit.eq}
A_{{\rm H}\alpha}=b+c \log r,
\end{equation}
with $r$ in kpc. 
We fit the dust profiles derived using
the {Calzetti} {et~al.} (2000) attenuation
curve, for consistency with previous work 
(e.g. {Momcheva} {et~al.} 2013; {Dom{\'{\i}}nguez} {et~al.} 2013; {Price} {et~al.} 2014) and
because this curve gives the best agreement with the
total UV+IR star formation rates (see Sect.\ 4).
For the three mass bins we find
$b=[0.0, 0.46, 1.10]$ and $c=[0.0, -0.90, -2.40]$. We set
$b$ and $c$ to zero in the lowest mass bin as the formal best
fit is (slightly) negative. 
These fits are shown by the thick dashed lines
in Fig.\,\ref{fig:ahaprofs}. 
While the observed
profiles show hints of more complex
structure, perhaps even
separate bulge and disk components,
higher order fits are not justified given the low
signal-to-noise ratio and the spatial resolution of the data.

We parameterize the change of the slope and normalization with 
\m\ by fitting linear functions to the coefficients $b(M)$
and $c(M)$. To good approximation ($\pm 0.05$ mag), we find
$b = 0.9 + 1.0 \log M_{10}$  and $c = -1.9 -2.2 \log M_{10}$,
with $M_{10}$ the stellar mass in units of $10^{10}$\,M$_{\odot}$.
Over a decade in mass, the attenuation at 2 kpc increases by
$\sim 0.5$ magnitudes, and the attenuation at 0.5 kpc increases by
$\sim 1.5$ magnitudes.  The gradients are consistent with the
galaxy-integrated measurements of {Price} {et~al.} (2014).


\section{Dust-Corrected Radial Profiles of Star Formation}
\begin{figure*}
\includegraphics[width=\textwidth]{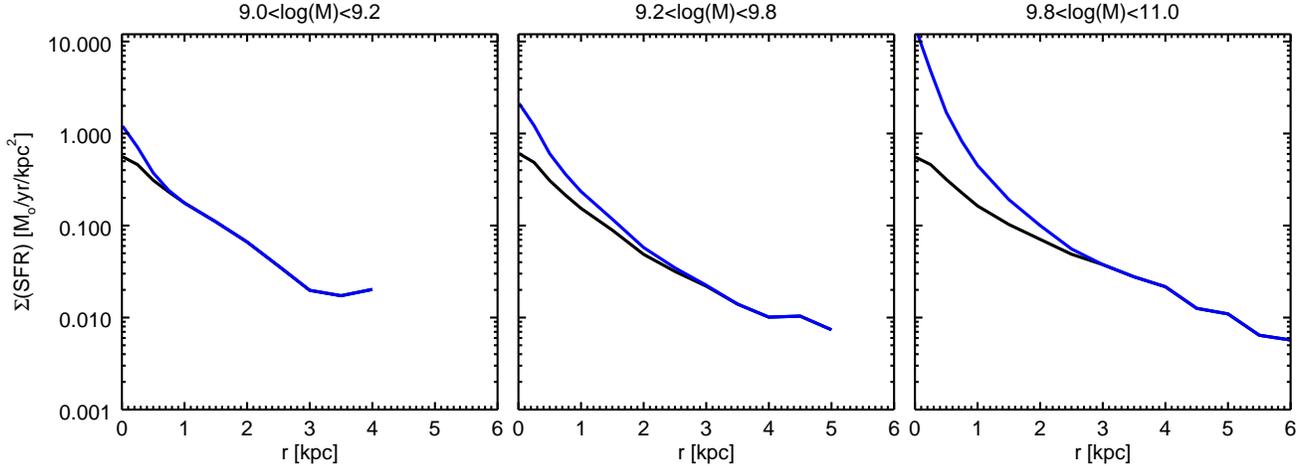}
\caption{Radial profiles of SFR determined directly from \ha\ emission 
are shown in black, radial profiles of SFR corrected for
dust using radial Balmer decrement profiles are shown in blue. 
The excess SFR over an exponential toward the center of high mass galaxies
suggests we are witnessing the growth of galactic bulges through in-situ 
star formation. 
\label{fig:dustcorrprofs}
}
\end{figure*}
With the radial profiles of \ha\ and \a, we derive radial profiles
of star formation corrected for dust. 
Star formation rates are computed from \ha\ emission 
using {Kennicutt} (1998) and a {Chabrier} (2003) initial mass function: 
\begin{equation}
SFR(H\alpha)=4.68\times10^{-42}L_{H\alpha}
\end{equation}
Star formation rates are corrected for the dust attenuation toward HII 
regions (\a) using:
\begin{equation}
SFR(dust\,corrected)=SFR(H\alpha)\times{\rm exp}({A_{H\alpha}/1.086}),
\end{equation}
with $A_{{\rm H}\alpha}$ given by Eq.\ \ref{fit.eq}. 

Figure\,\ref{fig:dustcorrprofs} shows the radial SFR profiles
corrected for dust.
As a test of our procedure,
we integrate these dust-corrected SFR profiles out to $r=2r_e$
(with $r_e$ the half light radius of the \f\ emission), 
and compare them to the average UV+IR star formation rates
of the same galaxies
(see  {Whitaker} {et~al.} 2014; {Momcheva} {et~al.} 2015). 
In the three mass bins, we find that the total H$\alpha$-derived
star formation rates are 2.4, 3.3, and 8.6 respectively. The ratios
of the H$\alpha$ star formation rates to the UV+IR star formation
rates are  1.3, 1.2, and 0.7 respectively. 
These ratios are similar when using the {Whitaker} {et~al.} (2014) 
UV+IR SFR measurements for the median masses in the bins. 
We conclude that the dust corrections are
reasonable to within the systematic uncertainty
of our UV+IR star formation rates (see e.g. {Utomo} {et~al.} 2014).%
\footnote[4]{The small discrepancy in the highest mass bin could 
be due to star formation with very high optical depth.}

Particularly in our highest mass bin
the implied central SFRs are much higher than prior
to the dust correction: the total star
formation in the central $r<1$\,kpc is $5.6\,{\rm M_\odot}/{\rm yr}$
in the highest mass bin, whereas it was $0.9\,{\rm M_\odot}/{\rm yr}$
before correction. Approximately half of the total star formation rate takes
place inside 1\,kpc in the highest mass bin.

\section{Discussion}

Using the WFC3 grism to map the distribution of \ha\ and \hb\ emission, we constructed
the first spatially-resolved maps of the Balmer decrement at $z>1$. These measurements
provide stringent constraints on the radial gradients in the dust attenuation toward
star forming regions of galaxies, allowing us to derive dust-corrected radial distributions
of star formation.  We find that the dust attenuation is small ($<0.5$\,mag) at all radii
in galaxies with $M<10^{10}$\,M$_{\odot}$. Galaxies with higher masses have significant
dust attenuation toward their centers.
The immediate implication is that the central ($r \lesssim 2$\,kpc)
observed \ha\ emission of high mass galaxies
should not be directly converted to star formation, and the central surface brightness
should not be directly converted to stellar mass density.

A potential concern in this analysis is that by stacking small galaxies with high 
attenuation and large galaxies with low attenuation, we could infer
a radial dust gradient where on an individual galaxy basis, there
is none. 
To test this, we remove all compact galaxies with sizes more than 
0.1 and 0.3\,dex below the size-mass relation from the stack. 
In both cases, the qualitative trends remain unchanged meaning the 
gradients are real, not a byproduct of stacking a 
heterogeneous sample. 
Another concern is that weighting galaxies by their \f\ flux biases the 
stacks toward galaxies with high \ha\ and \hb\ equivalent widths. 
If galaxies with high equivalent widths have preferentially low 
dust attenuation, this analysis could underestimate the 
true dust attenuation at the median mass of the stacks. 

A straightforward interpretation of our results is that we see the in-situ
building of bulges in massive galaxies at $z=1.4$. However, a key question is
whether this central star formation accounts for a significant
fraction of the stars in the central kpc of present-day
galaxies. The total amount of stars that are formed in the central
kpc  can be approximated by
\begin{equation}
M \sim w\,{\rm SFR}\,\tau,
\end{equation}
with $w \approx 0.6$ a correction
for mass loss due to stellar winds (see e.g. Bruzual \& Charlot 2003)
 and $\tau$ the duration of the star formation.
For $\tau \sim 2$\,Gyr (that is, assuming that the current star formation
rate is maintained until $z\sim 0.8$), we find that the total
mass that is added is $\sim 7\times 10^9$\,\msun, or $\sim 1/3$ of the
total mass of the galaxy. If $\tau\lesssim 1$\,Gyr the added mass is much
less significant.
We note that this does not necessarily imply that galaxies are growing 
``outside-in": this depends on the radial $M_*/L$ profiles
({Szomoru} {et~al.} 2013) which are driven by gradients in dust and age.
This could be tested by deriving mass profiles from the 
high spatial resolution CANDELS multi-color imaging. 

In summary, we infer that it is possible that a central mass concentration
is built up through in-situ star formation in the highest
mass galaxies, but only if the radial distribution
of star formation observed at $z\sim 1.4$ is
sustained over several Gyr.
This can be tested
with spatially-resolved absorption
maps at lower
redshift, which can be created
using adaptive optics or the G102 grism on the WFC3
camera. Furthermore, high resolution imaging of the molecular
gas and of the continuum dust emission with ALMA can provide
direct information on the presence of large amounts of dust
in the centers of massive galaxies at these redshifts.
Finally, with future facilities such as JWST
our initial study of averaged spatially-resolved Balmer
decrements at moderate redshifts can be expanded to higher
redshifts and individual galaxies.



\end{document}